\documentclass[prd,showpacs,letterpaper,twocolumn,nofootinbib,longbibliography,floatfix,superscriptaddress]{revtex4-1}%
\usepackage{graphicx}
\usepackage{bm}
\usepackage{epsf}
\usepackage{rotating}
\usepackage{epsfig,graphics,rotate,color}
\usepackage{wrapfig}
\usepackage{amssymb}
\usepackage{amsmath}
\usepackage{amsfonts}
\usepackage{multirow}
\usepackage{braket}
\usepackage[utf8]{inputenc}
\usepackage[dvipsnames]{xcolor}
\usepackage[colorlinks=true,citecolor=blue,linkcolor=blue]{hyperref}
\usepackage{cleveref}
\usepackage{mathbbol}
\usepackage{tabularx}
\usepackage{listings}
\usepackage{placeins}
\usepackage{booktabs}
\usepackage{accents}

\begin{document}
\newcommand\barparen[1]{\overset{(---)}{#1}}
\newcommand\brobor{\smash[b]{\raisebox{0.6\height}{\scalebox{0.5}{\bf{\tiny(}}}{\mkern-1.5mu\scriptstyle-\mkern-1.5mu}\raisebox{0.6\height}{\scalebox{0.5}{\tiny)}}}}

\title{Combining Sterile Neutrino Fits to Short Baseline Data with IceCube Data}
\author{M.H. Moulai}
\email{marjon@mit.edu}
\affiliation{Dept.~of Physics, Massachusetts Institute of Technology, Cambridge, MA 02139, USA}
\author{C.A. Arg\"uelles}
\affiliation{Dept.~of Physics, Massachusetts Institute of Technology, Cambridge, MA 02139, USA}
\author{G.H. Collin}
\affiliation{Dept.~of Physics, Massachusetts Institute of Technology, Cambridge, MA 02139, USA}
\affiliation{Institute for Data, Systems, and Society, Massachusetts Institute of Technology, Cambridge, MA 02139, USA}
\author{J.M. Conrad}
\affiliation{Dept.~of Physics, Massachusetts Institute of Technology, Cambridge, MA 02139, USA}
\author{A. Diaz}
\affiliation{Dept.~of Physics, Massachusetts Institute of Technology, Cambridge, MA 02139, USA}
\author{M.H. Shaevitz}
\affiliation{Dept.~of Physics, Columbia University, New York, NY, 10027, USA}

\begin{abstract}
Recent global fits to short-baseline neutrino oscillation data have been performed finding preference for a sterile neutrino solution (3+1) over null.
In the most recent iteration, it was pointed out that an unstable sterile neutrino (3+1+decay) may be a better description of the data.
This is due to the fact that this model significantly reduces the tension between appearance and disappearance datasets.
In this work, we add a one-year IceCube dataset to the global fit obtaining new results for the standard 3+1 and 3+1+decay sterile neutrino scenarios.
We find that the 3+1+decay model provides a better fit than the 3+1, even in the presence of IceCube, with reduced appearance to disappearance tension. The 3+1+decay model is a 5.4$\sigma$ improvement over the null hypothesis and a 2.8$\sigma$ improvement over the standard 3+1 model.
\end{abstract}


\maketitle

%---------------------------------------------------------
\section{Introduction\label{sec:intro}}

Over the last decades, anomalies observed in short-baseline (SBL) neutrino oscillation experiments~\cite{Aguilar:2001ty,MB2018} have motivated global fits that expand from a model containing three neutrinos to one that also includes a fourth mass state and a non-weakly interacting flavor state~\cite{Collin:2016aqd,Dentler:2018sju,Diaz:2019fwt,Boser:2019rta}.
Hence, these so-called ``3+1 models'' extend the neutral-lepton mixing matrix to be $4\times4$, introducing three new independent mixing elements: $U_{e 4}$, $U_{\mu 4}$, and $U_{\tau 4}$.
The preferred regions of $|U_{e 4}|$, $|U_{\mu 4}|$ and the new squared mass difference, $\Delta m^2_{41}$, have been reported in a recent global fit of the short-baseline data to a 3+1 model ~\cite{Diaz:2019fwt}.

A 3+1 model can simultaneously explain anomalies observed in ``appearance experiments'' ($\overset{\textbf{\fontsize{1pt}{1pt}\selectfont(--)}}{\nu}\hspace{-2pt}_\mu \rightarrow \overset{\textbf{\fontsize{1pt}{1pt}\selectfont(--)}}{\nu}\hspace{-2pt}_e$)~\cite{Aguilar:2001ty,MB2018} and in ``disappearance experiments'' ($\overset{\textbf{\fontsize{1pt}{1pt}\selectfont(--)}}{\nu}\hspace{-2pt}_e \rightarrow \overset{\textbf{\fontsize{1pt}{1pt}\selectfont(--)}}{\nu}\hspace{-2pt}_e$) primarily due to reactor experiments~\cite{reactor1,DANSS,NEOS}.
However, in a 3+1 model, the combined anomalies also predict a signal in ``$\overset{\textbf{\fontsize{1pt}{1pt}\selectfont(--)}}{\nu}\hspace{-2pt}_\mu$ disappearance experiments.''
For oscillations in vacuum, observed through charged-current scattering, the elements $|U_{e4}|^2$ and $|U_{\mu 4}|^2$, define three mixing angles that characterize the amplitude of $\nu_e$ disappearance, $\nu_\mu$ disappearance and $\nu_\mu \rightarrow \nu_e$ appearance, respectively:   
\begin{equation}
\begin{aligned}
\sin^22\theta_{ee} &= 4(1-|U_{e4}|^2)|U_{e4}|^2, \\
\sin^22\theta_{\mu \mu } &= 4(1-|U_{\mu 4}|^2)|U_{\mu 4}|^2, \\
\sin^2 2\theta_{\mu e} &= 4|U_{e4}|^2 |U_{\mu4}|^2. 
\label{eq:mixing_amplitudes}
\end{aligned}
\end{equation}
Thus, the mixing angles measured in the three types of searches are
not independent. 
In fact, for small mixing elements, they are approximately related in the following way:
\begin{equation}
\sin^22\theta_{ee}\sin^22\theta_{\mu \mu } \approx 4 \sin^2 2\theta_{\mu e}. 
\label{eq:unitarity_relationship}
\end{equation}

The squared-mass splitting must also be consistent for all three types of oscillations.
The probability that all SBL data, observed anomalies and constraints from null observations, is a likely realization of the 3+1 model is found to be extremely small~\cite{Dentler:2018sju}. 
This statement follows from a recent parameter goodness of fit test~\cite{Maltoni:2003cu} (``PG test'') performed on the global data, which quantifies the tension as a disagreement at the  $4.5\sigma$~\cite{Dentler:2018sju} level.   

If the anomalies are due to new physics, the underlying model may be more complicated than 3+1.
We have shown that adding another additional mass state, a model known as ``3+2,'' does not reduce the tension~\cite{Diaz:2019fwt}.
Other modifications are also possible~\cite{Gninenko:2009ks,Gninenko:2010pr,Masip:2012ke,Radionov:2013mca,,Blennow:2016jkn,Ballett:2018ynz,Bertuzzo:2018itn,Arguelles:2018mtc,Liao:2018mbg,Denton:2018dqq,Ballett:2019pyw, Fischer:2019fbw} and we have developed a 3+1 model that incorporates decay of the largest mass state, $m_4$~\cite{Moss:2017pur,Diaz:2019fwt}.
In fact, this category of 3+1 models was first considered as an explanation of the LSND observation in~\cite{PalomaresRuiz:2005vf} and was later realized to weaken the muon-neutrino disappearance constraints in~\cite{Moss:2017pur}.
More recently, it has been suggested that if one considers visible decay with one or more decay daughters being detectable, the low-energy excess of MiniBooNE can be well-described by this category of models~\cite{ivan_esteban_2019_3509890,Dentler:2019dhz,deGouvea:2019qre}.
In the case of invisible decay, where the heavy neutrino has two decay daughters that are undetectable, this introduces only one additional parameter beyond the 3+1 model, the lifetime of the heavy neutrino, $\tau$. However, it also introduces two new, negligible-mass particles, the decay daughters, into the phenomenology.
This model reduces the tension, measured by the parameter goodness of fit test, significantly to $3.2\sigma$~\cite{Diaz:2019fwt}.
While this still suggests poor agreement, it points to the possibility of additional effects in the short-baseline sample.

The global fits described above were limited to experiments studying vacuum oscillations at $L/E \sim 1$ to 10 eV$^2$. 
However, the IceCube experiment offers a relevant dataset that is very different from these short-baseline experiments. 
This experiment searches for a resonance signature from sterile-induced matter-effects in upward-going antineutrinos, which have traveled through the Earth~\cite{Nunokawa:2003ep,TheIceCube:2016oqi}.
The oscillation amplitude is no longer given by the vacuum-oscillation mixing angle relations, Eqs.~\ref{eq:mixing_amplitudes}, and instead takes on a complicated form, where $\sin^22\theta_{\mu\mu}$, the amplitude of the disappearance, depends on $|U_{\mu4}|$ and also $|U_{\tau 4}|$; see Ref.~\cite{Esmaili:2013fva,Lindner:2015iaa,Akhmedov:2016hcb,Blennow:2018hto} for a detailed discussion.
Thus, IceCube brings additional information to the global fits, since the vacuum oscillation results are insensitive to $|U_{\tau 4}|$~\cite{Collin:2016aqd}.

A 3+1 model has been explored for a combined short-baseline and IceCube global fit for the first time in~\cite{Collin:2016aqd} and later in~\cite{Dentler:2018sju}.
In this paper, we will follow the same procedure of~\cite{Collin:2016aqd} to include IceCube in the latest global fits given in~\cite{Diaz:2019fwt}, and we will expand the result to include 3+1+decay.

\section{IceCube}

\begin{figure}[ht]
  \includegraphics[width=\columnwidth]{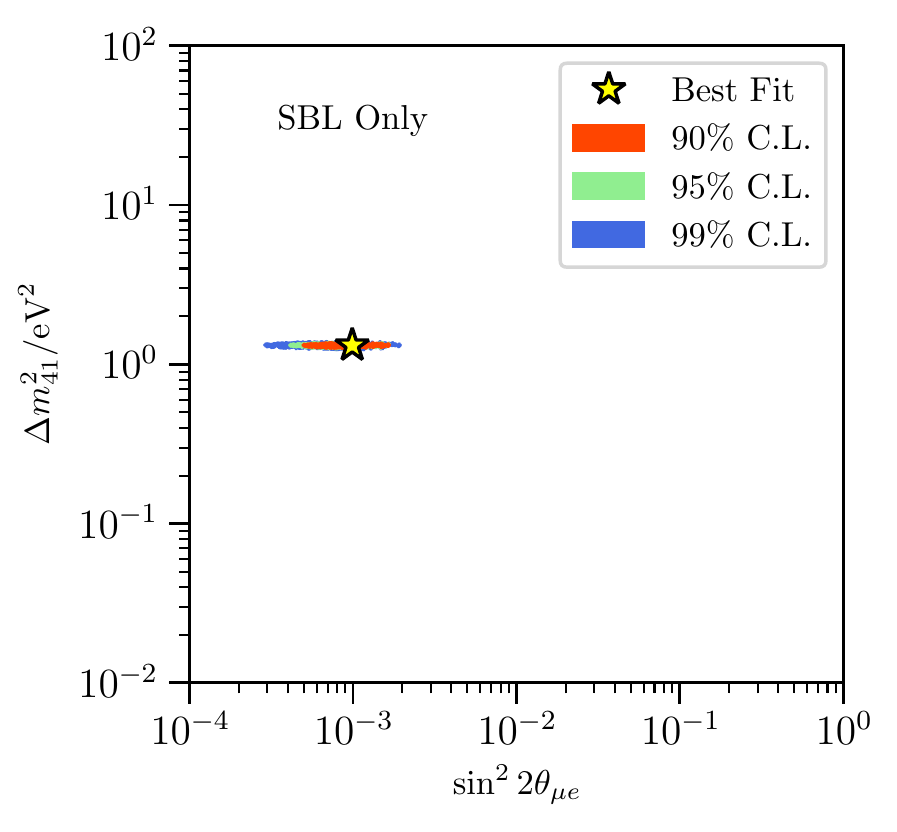} \\
  \includegraphics[width=\columnwidth]{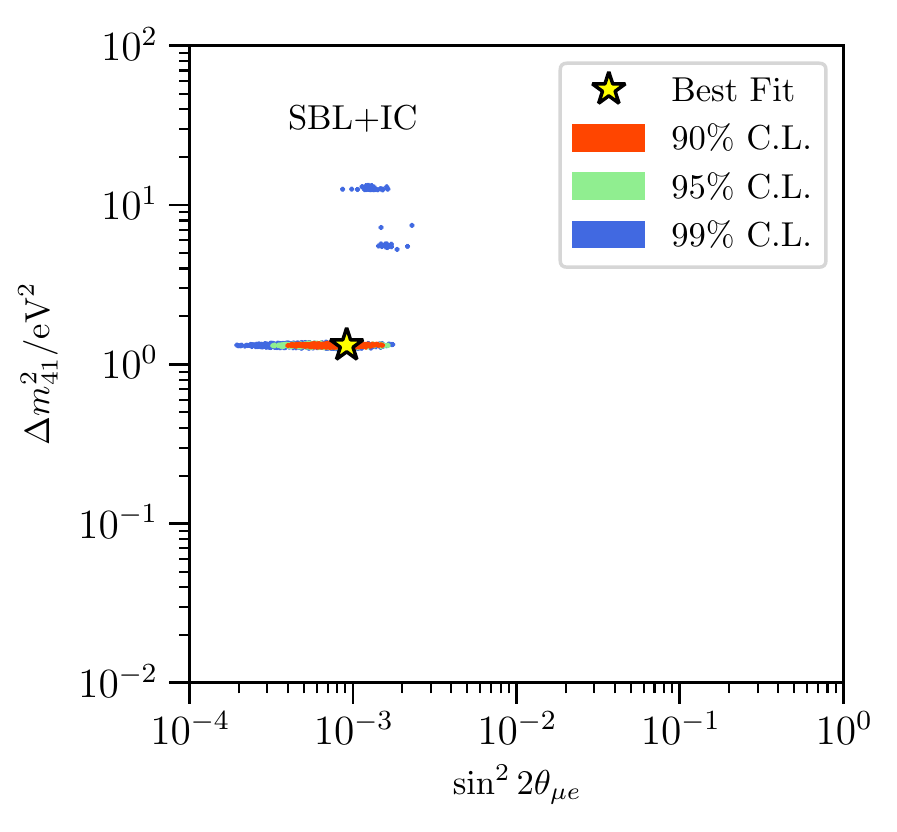}
  \caption{\textbf{\textit{The 3+1 allowed regions for fits to SBL experiments with and without IceCube.}} Top panel shows the allowed regions when only considering the SBL experiments, while the bottom panel shows them with IceCube incorporated. Frequentist confidence regions are shown at the 90\% (red), 95\% (green), and 99\% (blue) confidence levels.}
  \label{fig:gl_sbl_allowed}
\end{figure}

\begin{figure*}[ht]
  \includegraphics[width=\textwidth]{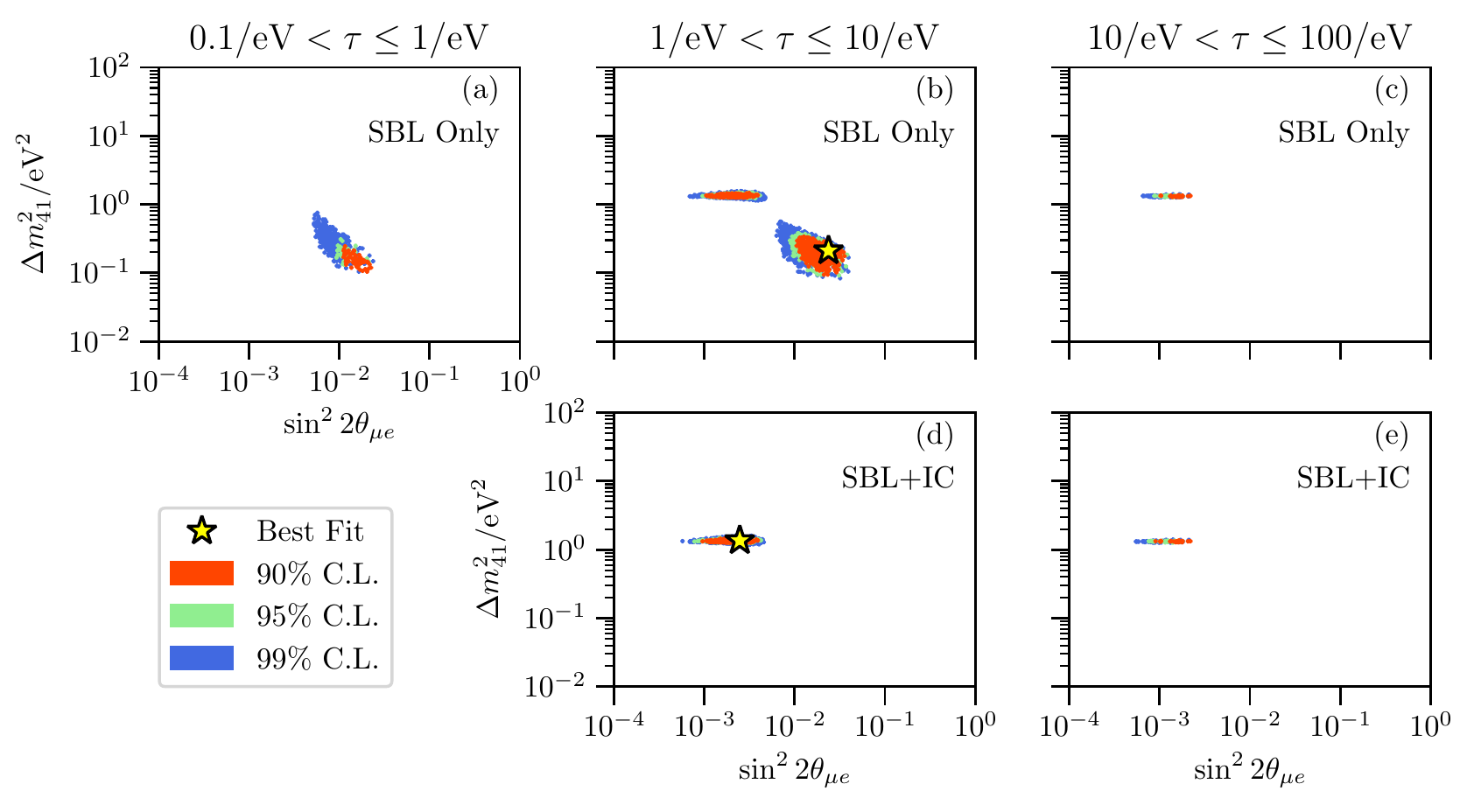}
  \caption{\textbf{\textit{The 3+1+decay frequentist allowed regions.}} The top row is the result of the fit using only the SBL experiments, while the bottom row includes IceCube. Each vertical column corresponds to a given range of $\nu_4$ lifetime, $\tau$. The smallest range of lifetimes in the SBL+IC is not shown as it contains no points. The colors of the points indicate 90\% (red), 95\% (green), and 99\% (blue) confidence levels.}
  \label{fig:gl_decay_sbl+ic_allowed}
\end{figure*}

The IceCube Neutrino Observatory is located in the Antarctic continent close to the geographic South Pole~\cite{Aartsen:2016nxy}.
IceCube is a gigaton-scale ice-Cherenkov detector made out of arrays of photomultiplier tubes encapsulated in pressure-resisting vessels buried in the clear Antartic glaciar ice~\cite{Aartsen:2013rt}.
IceCube has measured the atmospheric neutrino spectrum from ~10~GeV~\cite{Aartsen:2017nmd}, using a denser inner part of the detector called DeepCore~\cite{Collaboration:2011ym}, to 100 TeV~\cite{Aartsen:2017nbu}.
The atmospheric neutrino flux is dominated by the so-called conventional component, which is due to the decay of kaons, pions, and muons~\cite{Gaisser:2002jj}. 
At the lower energies, below 100~GeV, neutrinos from pion and muon decay dominate the neutrino flux and IceCube has used them to measure the atmospheric oscillation parameters~\cite{Aartsen:2017nmd}.
In this energy range, IceCube is also sensitive to eV-scale sterile neutrinos by looking for modifications on top of the predicted standard three-neutrino oscillation pattern~\cite{Aartsen:2017bap}; as similarly done by SuperKamiokande~\cite{Abe:2014gda}.
At TeV energies, neutrino oscillations driven by the known squared-mass differences turn off, as their oscillation length becomes much larger than the Earth's diameter.
It was pointed out in~\cite{Nunokawa:2003ep}, that at TeV energies matter effects will induce a large disappearance of muon-antineutrinos for squared-mass differences compatible with the LSND anomaly, due to effects previously studied in broader contexts in~\cite{Akhmedov:1988kd,Krastev:1989ix,Chizhov:1998ug,Chizhov:1999az,Akhmedov:1999va}.
IceCube has performed a search for sterile neutrinos using one year of data~\cite{TheIceCube:2016oqi}. 
This analysis used a high-purity muon-neutrino event selection designed to search for an astrophysical component in the northern sky~\cite{Aartsen:2015rwa}.
The analysis was performed with neutrinos between 400~GeV and 20~TeV in energy; this range was chosen to avoid contamination from high-energy astrophysical neutrinos  and to minimize uncertainties in local ice properties at lower energies.

The IceCube collaboration provided a data release associated with this analysis~\cite{TheIceCube:2016oqi}.
This data release includes over 20,000 atmospheric $\nu_\mu$ and $\bar{\nu}_\mu$ events, as well as the Monte Carlo that was used.
In this work, we use the analysis tools we developed in~\cite{Collin:2016aqd,Moss:2017pur}, which use the nuSQuIDS package to calculate neutrino oscillation probabilities~\cite{Delgado:2014kpa,nusquids}.
In our analysis of the IceCube data we consider two sources of uncertainties.
One source of uncertainty is the atmospheric neutrino flux, which we parameterize by means of nuisance parameters.
We consider: the overall normalization of the atmospheric flux, the cosmic-ray slope, the uncertainty in the atmospheric density, the ratio of kaon-to-pion production yields, and the ratio of neutrino-to-antineutrino production.
The second source of uncertainty is the detector. As discussed in~\cite{2015PhDT........94D,Jones:2015bya}, in this analysis we restrict ourselves to the leading detector systematic, namely the overall efficiency of the IceCube modules.
Following~\cite{TheIceCube:2016oqi}, the IceCube likelihood can be written as
\begin{equation}
\begin{aligned}
    \ln \mathcal{L} = \min_\eta \bigg( 
    &\sum_{i=1}^{N_{\rm{bins}}}
    \big[ x_i \ln \mu_i (\vec{\theta}, \vec{\eta}) - \mu_i (\vec{\theta}, \vec{\eta}) - \ln x_i! \big] \\
    &+ \frac{1}{2} \sum_{\eta} \frac{(\eta - \bar{\eta})}{\sigma_\eta^2} \bigg),
\end{aligned}
\end{equation}
where $x_i$ is the number of data events in the $i^{th}$ bin; $\mu_i$ is the Monte Carlo expectation for the number of events in the $i^{th}$ bin, assuming sterile neutrino parameters $\vec{\theta}$ and nuisance parameters $\vec{\eta}$; and each nuisance parameter, $\eta$, has a Gaussian constraints of mean, $\bar{\eta}$, and standard deviation, $\sigma_\eta$.
We use the systematic treatment and analysis framework from~\cite{Moss:2017pur} to calculate the IceCube likelihood.

For both the 3+1 and 3+1+decay global fits, the IceCube likelihood was calculated for a randomly-selected subset of parameter-set points from the corresponding Markov Chain Monte Carlo from~\cite{Diaz:2019fwt}.
This global analysis used experiments where vacuum oscillation probabilities are valid and, in the stable 3+1 model, those were parameterized in terms of $|U_{e 4}|$, $|U_{\mu 4}|$, and $\Delta m^2_{41}$.
The data sets used include all relevant muon-(anti)neutrino disappearance, electron-(anti)neutrino disappearance, and muon-to-electron (anti)neutrino appearance measurements; a list of the experiments used can be found in the Supplementary Material.
There are four notable exceptions: 1) the Daya Bay result~\cite{An:2016luf}, as it does not significantly impact the preferred best-fit region ($\Delta m_{41} > 1 {\rm eV^2}$)~\cite{Collin:2016aqd}; 2) the reactor anomaly due to the large uncertainties in the reactor flux modeling made manifest by the so-called 5~MeV reactor bump~\cite{Huber:2016fkt,Dentler:2017tkw}; the low-energy atmospheric neutrino results from SuperKamiokande~\cite{Abe:2014gda} and IceCube/DeepCore~\cite{Aartsen:2017bap} as these constraints only reach $|U_{\mu 4}| > 0.2$ which are covered by other experiments in the mass-square region of interest; and 4) the recent MINOS+ two-detector fit~\cite{Adamson:2017uda,Adamson:2020jvo}, due to the lack of clear explanation of the origin of the sensitivity in the high-mass region and increased systematic model dependence, instead we use the traditional far-to-near ratio measurement by MINOS~\cite{MINOS:2016viw}. 
Each point is a unique combination of $|U_{e 4}|$, $|U_{\mu 4}|$, and $\Delta m^2_{41}$; $\tau$ is an additional parameter for the 3+1+decay scenario.
In this work we set $|U_{\tau 4}|$ to zero, as the short-baseline experiments are insensitive to it and this is a conservative choice in the case of IceCube~\cite{Collin:2016aqd}.
This downsampling is necessary because the IceCube likelihood calculation is computationally time-intensive.
We include the best-fit parameter-set points corresponding to 3+1 and 3+1+decay found in~\cite{Diaz:2019fwt}.
For the 3+1 analysis 49000 points were used, while for the 3+1+decay analysis 82000 points were used.

To convert the IceCube likelihood into a $\chi^2$, we calculate the log-likelihood ratio~\cite{Collin:2016aqd}
\begin{equation}
    \ln \mathcal{LR} (\vec{\theta}) = \ln \big( \mathcal{L}(\vec{\theta}) \big) - \ln \big( \mathcal{SP}\{(x_i)\} \big),
\end{equation}
where $\mathcal{SP}$ $\{(x_i)\}$ is the saturated Poisson likelihood and then assume Wilks' theorem; namely $\chi^2 = -2 \ln \mathcal{LR}$~\cite{wilks1938,Olive:2016xmw}.
To incorporate IceCube into the global fits, we add the IceCube $\chi^2$ to the $\chi^2$ from the SBL-only fits~\cite{Diaz:2019fwt}.

To determine the effect that IceCube data has on the tension for both the 3+1 and 3+1+decay models, we calculate the IceCube likelihood for a random downsampling of parameter points from the recent fits to only disappearance SBL experiments.
We convert the IceCube likelihood to a $\chi^2$ and add it to the SBL disappearance $\chi^2$. The fit to the appearance experiments remains unchanged.

\section{Results\label{Results}} 

In this section we summarize the result of incorporating IceCube into our recent SBL-only global fit.
Results for fitting a standard 3+1 model are discussed in Sec.~\ref{3+1 results}, while results for fitting a 3+1+decay model are shown in Sec.~\ref{3+1+Decay results}.
A summary of the $\chi^2$ values for various fits and additional figures for all analyses performed in this work are given in the Supplementary Material.

\subsection{3+1\label{3+1 results}} 

In this section we report the impact of adding IceCube to the short-baseline-only 3+1 fit. 
Fig.~\ref{fig:gl_sbl_allowed} shows the frequentist allowed regions without IceCube data (top) and with IceCube data incorporated (bottom), for a random downsampling of parameter points used in the recent global fit~\cite{Diaz:2019fwt}.
The allowed regions are shown in terms of $\Delta m^2_{41}$ and $\sin^2 2 \theta_{\mu e}$, given by Eq.~\ref{eq:mixing_amplitudes}, at 90\%, 95\%, and 99\% confidence levels.
The best-fit points are indicated in each panel with a star.
Before including IceCube data, the downsampled points produce allowed regions that are consistent with our previous result and are shown here for completeness.
Incorporating IceCube data pushes the island at $\Delta m^2_{41} \approx 1.3 \rm{eV}^2$ to slightly smaller $\sin^2 2 \theta_{\mu e}$, and creates new islands at higher masses at the 99\% confidence level.
A table of the best-fit parameters is given in Tab.~\ref{table:3_1_fit_results} and the best-fit $\Delta \chi^2$ is given in Tab.~\ref{table:short_chi2_table}. The obtained best-fit values on the 3+1 model are in tension with the MINOS+ results not included in this work for the reasons mentioned in Sec. II. The significance of the best-fit point of the 3+1 model compared to the null hypothesis, {\it i.e.} three neutrinos, is 4.9$\sigma$, slightly lower than it is without IceCube, 5.1$\sigma$.
We have performed the parameter goodness of fit test to quantify the tension in the dataset.
The significance of this tension with IceCube included is 4.8$\sigma$, which is slightly higher than the value without including IceCube, which is 4.5$\sigma$.
~
\begin{table}[ht]
\begin{center}
  \begin{tabular}{ | c | c  c  c  |  }
    \hline
    3+1 & $\Delta m_{41}^2$ (eV$^2$) & $|U_{e4}|$ & $|U_{\mu4}|$  \\ \hline
    \hline
    %SBL (RMP)    & 1.32  & 0.116 & 0.135 \\ \hline \hline
    SBL          & 1.32  & 0.116 & 0.135  \\ \hline
    IC           & 6.97 & -- & 0.155\\ \hline
    SBL+IC       & 1.32 & 0.116 & 0.131 \\
    \hline
  \end{tabular}
   \caption{\textbf{\textit{3+1 model best-fit parameters.}} Each row corresponds to a different dataset used. The columns give the best-fit value of the parameter for the listed dataset.}
  \label{table:3_1_fit_results}
 \end{center}
\end{table}

\subsection{3+1+Decay\label{3+1+Decay results}}

Global fit results for the 3+1+decay model are shown in Fig.~\ref{fig:gl_decay_sbl+ic_allowed}.
The top row of this figure shows the allowed regions without including IceCube, which are consistent with previous results, while the bottom row shows the allowed regions with IceCube incorporated.
The allowed regions without considering IceCube span three orders of magnitude in $\tau$, and are shown in three panels, each corresponding to the indicated range in $\tau$. The allowed regions without IceCube occur in two distinct regions in $\Delta m_{41}^2$.
IceCube data eliminates the lower squared-mass island, and collapses the allowed squared masses to a narrow range around $\Delta m_{41}^2 \approx 1.4~ \rm{eV}^2$.

A table of the best-fit parameters is given in Tab.~\ref{table:decay_fit_results} and the best-fit $\Delta \chi^2$ is given in Tab.~\ref{table:short_chi2_table}.
The significance of the best-fit point of the 3+1+decay model compared to the three neutrino hypothesis is 5.4$\sigma$, slightly lower than it is without IceCube, which is 5.6$\sigma$. Notably, the 3+1+decay model is preferred to the standard 3+1 model by 2.8$\sigma$.
Our global-fit combines three different oscillation channels and the 3+1+decay model introduces features in all three; see Supplementary Material for details.
In the reactor electron-antineutrino disappearance measurements, the best-fit decay reduces the oscillation features at low energies where the statistics are larger.
Similarly, in muon-neutrino disappearance measurements, it reduces the oscillation amplitude for baselines greater than the decay length.
In the appearance channel, for scales relevant for MiniBooNE, the appearance probability maintains the spectral shape but has increased normalization.
Tension in the fits remains, at 3.5$\sigma$ with IceCube, yet it is reduced by about 1.3$\sigma$ compared to the standard 3+1 model.
Finally, note that the inclusion of absolute reactor flux normalization could yield important information to distinguish between models.
The preferred parameters of the 3+1+decay increases the electron-antineutrino disappearance from approximately 3\% in the 3+1 best-fit point to 10\% for reactor neutrino scales of interest.
This is due to the fact that the estimated neutrino flux uncertainties range from 2\% to 6\% in the 1 to 6~MeV~\cite{Hayen:2018uyg}; note also that the baseline flux model predictions did not foresee the appearance of a deviation at the 10\% level known as the 5~MeV bump.
Given these uncertainties we have performed a normalization independent analysis; including a reliable reactor flux uncertainty estimation would impact our obtained best-fit point.
~
\begin{table}
\begin{center}
  \begin{tabular}{ | c | c  c  c  c | }
    \hline
    3+1+Decay & $\Delta m_{41}^2$ (eV$^2$) & $\tau$ (eV$^{-1}$) & $|U_{e4}|$ & $|U_{\mu4}|$ \\ \hline
    \hline
    %SBL (RMP)    & 0.21  & 1.96 & 0.428 & 0.180 \\ \hline \hline
    SBL          & 0.21  & 1.96 & 0.428 & 0.180 \\ \hline
    IC           & 6.22  & 8.11 &   --   & 0.124 \\ \hline
    SBL+IC       & 1.35  & 4.50 & 0.238 & 0.105 \\
    \hline
  \end{tabular}
   \caption{\textbf{\textit{3+1+Decay model best-fit parameters.}} Each row corresponds to a different dataset used. The columns give the best-fit value of the parameter for the listed dataset. For $\tau = 1~{\rm eV}^{-1}$, $c\tau = 0.2 ~ \mu m $.}
  \label{table:decay_fit_results}
 \end{center}
\end{table}

 \begin{table}
 \setlength{\tabcolsep}{0.5em}
 \begin{center}
   \begin{tabular}{ | c | c | c | }
     \hline
         & 3+1 & 3+1+Decay \\ \hline
     \hline
     ($\Delta \chi^2$/$\Delta$dof)$_{\rm{Null}}$ & 31.7 / 3 & 40.7 / 4\\ \hline
     ($\Delta \chi^2$/$\Delta$dof)$_{\rm{3+1}}$ & -- & 9.1 / 1\\
     \hline
   \end{tabular}
   \caption{\textbf{\textit{Comparison of best-fit $\chi^2$ values.}} The first row gives the difference in $\chi^2$ and degrees of freedom between the null hypothesis (no sterile neutrino) and the best-fit values from global fits to a 3+1 model and 3+1+decay model.
   The second row gives the difference in best-fit $\chi^2$ and degrees of freedom between a 3+1 model and a 3+1+decay model. The results in this table include IceCube data.}
   \label{table:short_chi2_table}
  \end{center}
 \end{table}

\section{Conclusion\label{sec:conclusion}}

We have studied the impact of adding one year of IceCube high-energy atmospheric data to our short-baseline light sterile neutrino global fits. We considered two models in this work: one where the heavy neutrino mass state is stable and one where it decays to invisible particles. We summarize our findings here:
\begin{itemize}
\item For the case of stable neutrino states, we find that the best-fit solution to the short-baseline data is only slightly changed; the largest change is in $|U_{\mu 4}|$ which goes from 0.135 to 0.131.
Adding IceCube data makes two new solutions appear at the 99\% C.L. at larger squared-mass differences; these are at approximately $10~{\rm eV}^2$ and $5~{\rm eV}^2$, with $\sin^22\theta_{\mu e}\sim 10^{-3}$.
Even though the IceCube analysis is a null-like result, the best-fit 3+1 point is still significantly preferred over the null hypothesis by a $\Delta \chi^2 = 32$ with three degrees of freedom. This corresponds to a $4.9\sigma$ rejection of the null model.
As expected, adding the IceCube dataset increases the tension between appearance and disappearance datasets which, measured using the parameter goodness of fit test, worsens from a p-value of $3.4\times 10^{-6}$ without IceCube to $8.1\times 10^{-7}$ with it.
\item For the model where the heaviest mass state is allowed to decay into invisible particles, the preferred values of the parameters have changed significantly.
The new best-fit point results in a squared-mass difference of $1.35~{\rm eV}^2$, a lifetime of $4.5~{\rm eV}^{-1}$, and mixing elements $|U_{e 4}| = 0.238$ and $|U_{\mu 4}| = 0.105$.
As in the case of a stable neutrino mass states, this scenario is preferred over the null three-neutrino hypothesis at high significance; $5.6 \sigma$ without IceCube and $5.4 \sigma$ with it.
This model was already preferred with respect to the 3+1 scenario at the $2.7\sigma$ level~\cite{Diaz:2019fwt}, but with the addition of the IceCube data this preference increases to the $2.8\sigma$ level.
Finally, the tension is slightly worse when adding the IceCube data in this model; the p-value for the parameter goodness of fit test decreases from $8.0 \times 10^{-4}$ to $2.7 \times 10^{-4}$. Nevertheless, this model is still a factor of $\sim300$ improvement in p-value with respect to the stable sterile neutrino scenario.
\end{itemize}

In conclusion, the 3+1+decay scenario best-fit parameters have been significantly changed with the addition of one year of IceCube data. The new best-fit solution is still an improvement over the 3+1 scenario since it improves the overall fit and reduces the tension between appearance and disappearance experiments. We expect that the upcoming eight year IceCube high-energy sterile analysis may significantly impact the light sterile neutrino interpretation of the short-baseline anomalies.

\section*{Acknowledgements}

MHM, CAA, JMC, and AD are supported by NSF grant PHY-1801996.
MHS is supported by NSF grant PHY-1707971.
We thank Austin Schneider for useful comments.

%\pagebreak
\bibliographystyle{apsrev}
\bibliography{fits}

%%%%%%%%%% supplemental materials %%%%%%%%%%
\pagebreak
\clearpage

%%%%%% SUPLEMENTARY MATERIAL STARTS HERE
%%%%%% SUPLEMENTARY MATERIAL STARTS HERE
%%%%%% SUPLEMENTARY MATERIAL STARTS HERE
%%%%%% SUPLEMENTARY MATERIAL STARTS HERE

\onecolumngrid
\appendix

\ifx \standalonesupplemental\undefined
\setcounter{page}{1}
\setcounter{figure}{0}
\setcounter{table}{0}
\setcounter{equation}{0}
\fi

\renewcommand{\thepage}{Supplemental Material-- S\arabic{page}}
\renewcommand{\figurename}{SUPPL. FIG.}
\renewcommand{\tablename}{SUPPL. TABLE}

\renewcommand{\theequation}{A\arabic{equation}}
\clearpage

\begin{center}
\textbf{\large Supplemental Material}
\end{center}

\section{Illustrating tension between appearance and disappearance}

\begin{figure}[ht]
  \includegraphics[width=0.45\textwidth]{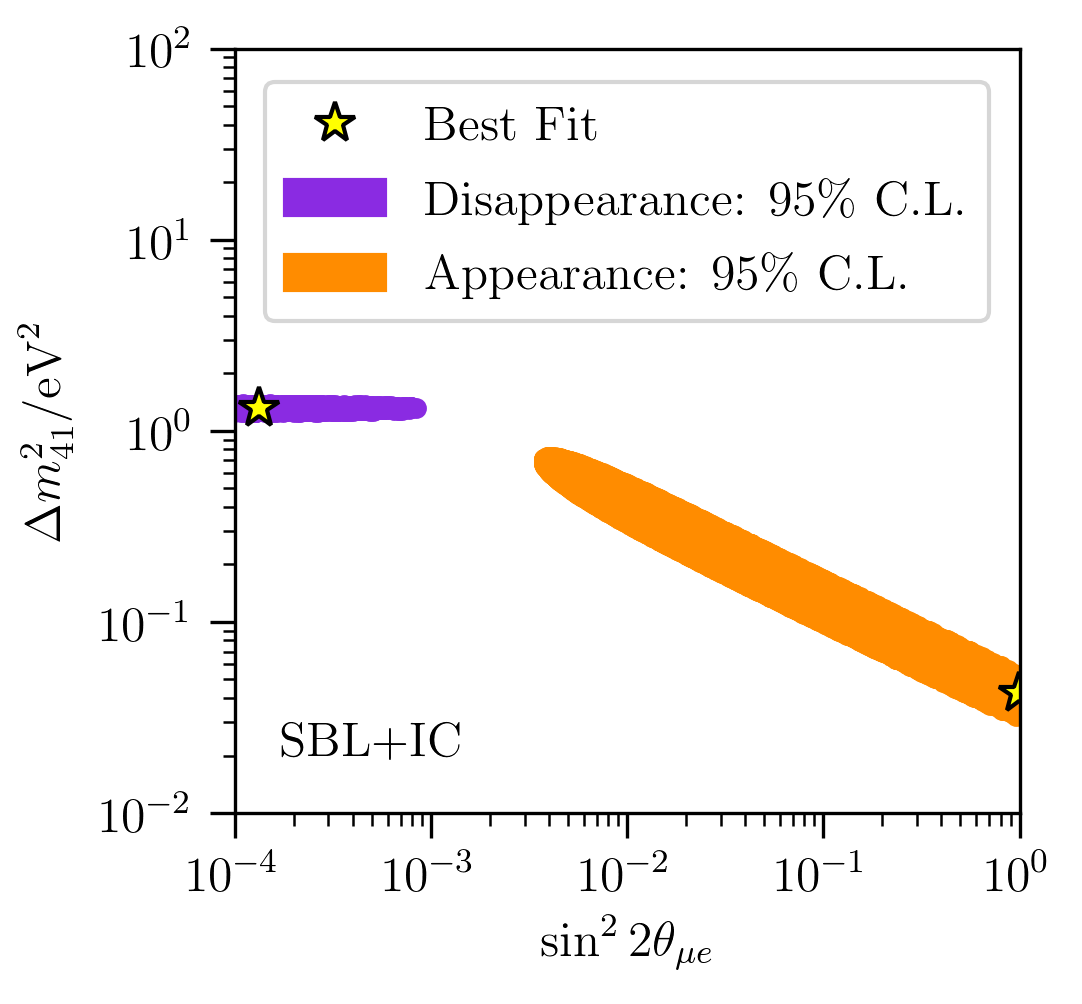}
  \caption{\textbf{\textit{Preferred regions for separate appearance and disappearance 3+1 fits.}} Frequentist 95\% C.L. regions for fits to a 3+1 model, performed separately to appearance-only data and disappearance-only data, are shown. There is no overlap, indicating tension in the fits.}
  \label{fig:tension_nodecay}
\end{figure}

\begin{figure*}[ht]
  \includegraphics[width=\textwidth]{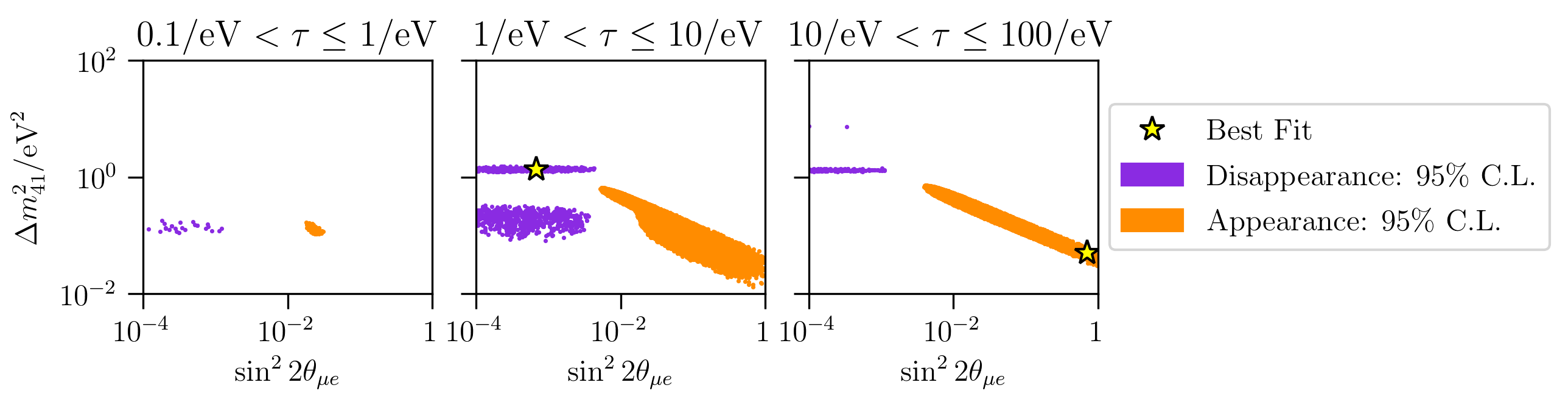}
  \caption{\textbf{\textit{Preferred regions for separate appearance and disappearance 3+1+decay fits.}} Frequentist 95\% C.L. allowed regions for fits to a 3+1+decay model, performed separately to appearance-only data and disappearance-only data, are shown. There is no overlap, indicating tension in the fits.}
  \label{fig:tension_nodecay_split}
\end{figure*}

Suppl.~Fig.~\ref{fig:tension_nodecay} and Suppl.~Fig.~\ref{fig:tension_nodecay_split} show frequentist allowed regions for separate fits to the appearance datasets and disappearance datasets at 95\% confidence level, on plots of $\Delta m^2_{41}$ versus $\sin^2 2 \theta_{\mu e}$. Suppl.~Fig.~\ref{fig:tension_nodecay} shows results for a 3+1 model, while Suppl.~Fig.~\ref{fig:tension_nodecay_split} shows results for a 3+1+decay model, separated into three lifetime decades. The disappearance datasets include IceCube. The best-fit points for each fit are indicated. In neither model is there overlap in the appearance and disappearance confidence intervals at 95\%, indicating tension in the fits.

\FloatBarrier
\pagebreak
\section{Preferred regions of 3+1+decay}

\begin{figure*}[hb]
  \includegraphics[width=0.8\textwidth]{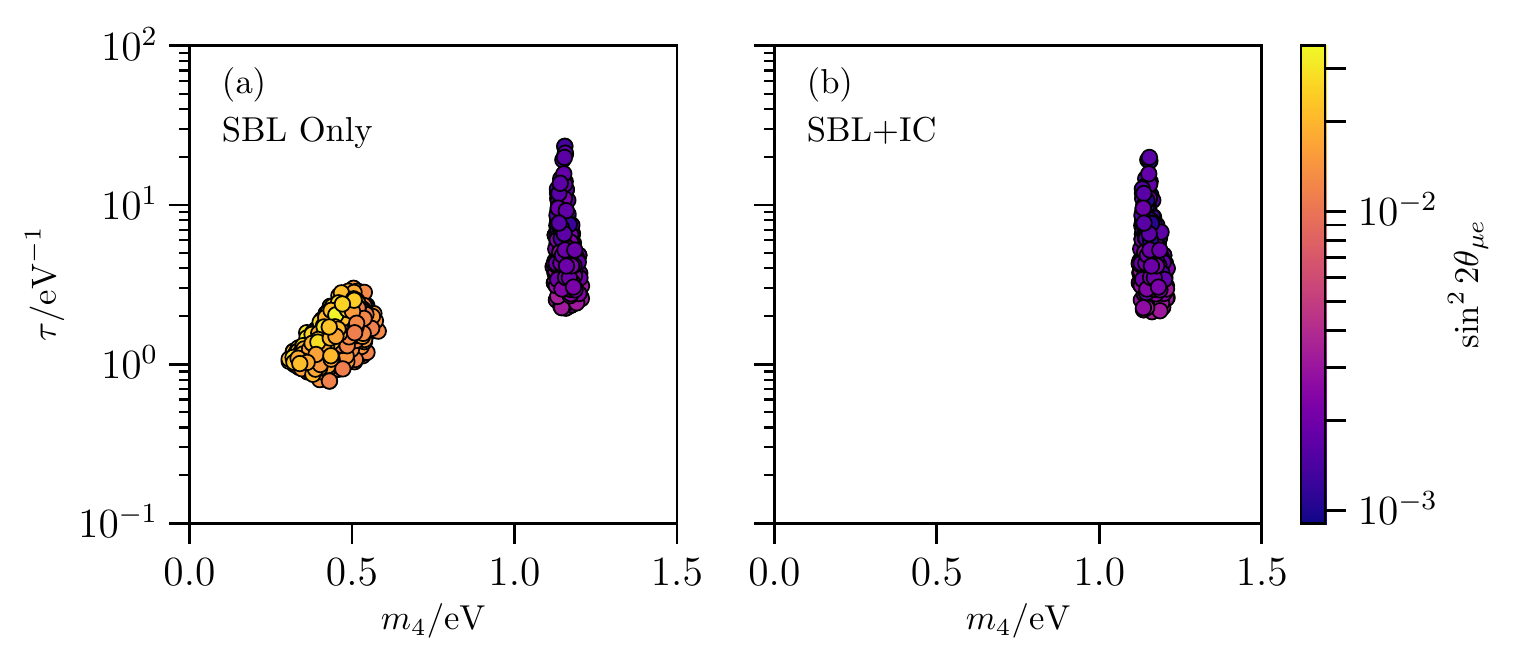}
  \caption{\textbf{\textit{3+1+decay model allowed regions at 90\% C.L.}} The left panel shows the allowed regions when considering only the short-baseline (SBL) dataset, while the right panel shows them with IceCube (IC) included. In both panels, the horizontal axis shows the new mass state mass, $m_4$, the vertical axis its lifetime, $\tau$, and the color scale the appearance amplitude. When considering only the short-baseline data two populations exist: one with large mixings and small masses ($\sin^22\theta_{\mu e} \sim 10^{-2}$ and $m_4 \sim 0.5 ~ {\rm eV}$) and another with smaller mixing and larger masses ($\sin^22\theta_{\mu e} \sim 10^{-3}$ and $m_4 \sim 1.25 ~ {\rm eV}$). When including IceCube, the large mass population is removed. For $\tau = 1~{\rm eV}^{-1}$, $c\tau = 0.2 ~ \mu m $.}
  \label{fig:mass_lifetime_angle}
\end{figure*}

\begin{figure}[ht]
  \includegraphics[width=0.6\textwidth]{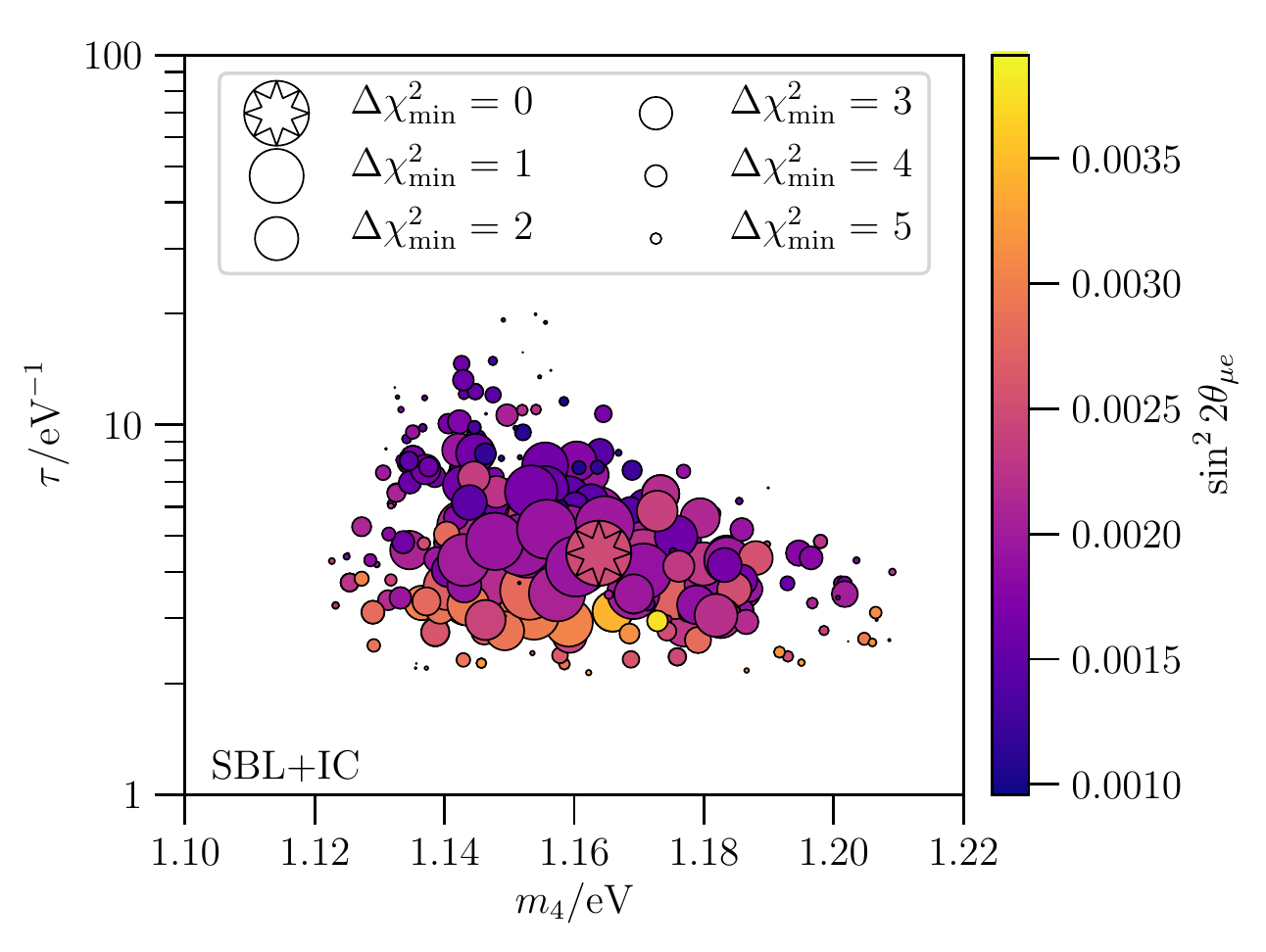}
  \caption{\textbf{\textit{3+1+decay model allowed regions using the short-baseline and IceCube data at 95\% C.L., shown with parameter point preference.}} In this figure the markers' size show the preference for a particular point, the horizontal axis shows the new mass state mass, $m_4$, the vertical axis its lifetime, $\tau$, and the color scale the mixing amplitude. The best-fit point is shown as the largest marker with the star symbol imprinted. Points with decreasing preference, with respect to the best-fit point, are drawn with smaller markers. For $\tau = 1~{\rm eV}^{-1}$, $c\tau = 0.2 ~ \mu m $.}
  \label{fig:decay_fit_2dof}
\end{figure}

Suppl.~Fig.~\ref{fig:mass_lifetime_angle} and Suppl.~Fig.~\ref{fig:decay_fit_2dof} show the global-fit result for a 3+1+decay model at 95\% C.L. as a function of the three relevant parameters: lifetime ($\tau$), heavy neutrino mass ($m_4$), and appearance amplitude ($\sin^22\theta_{\mu e}$). Suppl.~Fig.~\ref{fig:mass_lifetime_angle} shows the result for short-baseline experiments only (left) and IceCube incorporated (right). Suppl.~Fig.~\ref{fig:decay_fit_2dof} shows the result including IceCube, where the preference for each parameter point is indicated by the marker size.

\FloatBarrier
\pagebreak
\section{List of experiments included}

The data sets used in this work include neutrino and antineutrino sets; they are:
\begin{itemize}
    \item \textbf{Muon-to-electron neutrino appearance ($\nu_\mu \to \nu_e$)}: MiniBooNE (BNB)~\cite{Aguilar-Arevalo:2018gpe}, MiniBooNE (NuMI)~\cite{Adamson:2008qj}, NOMAD~\cite{Astier:2003gs}, LSND~\cite{Athanassopoulos:1997pv}, and  KARMEN~\cite{Armbruster:2002mp}.
    \item \textbf{Muon-neutrino disappearance ($\nu_\mu \to \nu_\mu$)}: SciBooNE/MiniBooNE~\cite{Mahn:2011ea}, CCFR~\cite{Stockdale:1984cg}, CDHS~\cite{Dydak:1983zq},
    MINOS~\cite{MINOS:2016viw}, and IceCube~\cite{TheIceCube:2016oqi}.
    \item \textbf{Electron-neutrino disappearance ($\nu_e \to \nu_e$)}: KARMEN/LSND cross section~\cite{Conrad:2011ce}, Bugey~\cite{Declais:1994su}, NEOS~\cite{Ko:2016owz}, DANSS~\cite{Alekseev:2018efk}, SAGE~\cite{SAGE}, GALLEX~\cite{Gallex}, and PROSPECT~\cite{Ashenfelter:2018iov}.
\end{itemize}

%\FloatBarrier
%\pagebreak
\section{Table of $\chi^2$}

Suppl.~Tbl.~\ref{table:chi2_summary} reports the $\chi^2$ obtained for the two different models considered in this work. 
For each model, we report the best-fit $\chi^2$ for the short-baseline dataset only and this set with IceCube included as different columns.
We also report the $\chi^2$ obtained when separating the data into appearance and disappearance experiments in order to evaluate the parameter goodness of fit.

\begin{table*}[ht]
\setlength{\tabcolsep}{0.5em}
  \begin{tabular}{ | c | c c | c  c | c c |}
    \hline
    Fit type           &3$\nu$ (Null) &3$\nu$ (Null)& 3+1   & 3+1     & 3+1+Decay & 3+1+Decay \\ 
    Dataset        & SBL & SBL+IC & SBL   & SBL+IC  & SBL       & SBL+IC \\ \hline
    Best fit           &&&       &       &       &  \\
    $\chi^2 / \rm{d.o.f.}$ &492.7 / 509 & 672.5 / 718 & 458.4 / 506 & 640.8 / 715 & 449.9 / 505 & 631.84 / 714 \\
    %d.o.f.:               &509&718  & 506   & 715   & 505   & 715\\
    p-value            &0.69&0.89& 0.94 & 0.98 & 0.96 & 0.99\\ \hline
    Null vs. Sterile   &&&       &       &       &  \\
    $\Delta\chi^2/\Delta\rm{d.o.f.}$&&& 34.3 / 3 & 31.7 / 3 & 42.8 / 4 & 40.7 / 4 \\
    %$\Delta d.o.f.$        &&& 3    & 3     &  4    &  4\\
    p-value:            &&& 1.7E-07& 6.0E-07& 1.2E-8 & 3.1E-8   \\
    $N \sigma$          &&& 5.1  &  4.9  &  5.6 &  5.4 \\ \hline
    3+1 vs. 3+1+decay   &&&      &       &       &  \\
    $\Delta\chi^2/\Delta\rm{d.o.f.}$&&&&   &  8.5 / 1 & 9.0 / 1 \\
   % $\Delta d.o.f.$        &&&       &     &  1  &  1\\
    p-value:            &&&       &      & 0.0036 & 0.0027   \\
    $N \sigma$          &&&       &      & 2.7 &  2.8 \\ \hline
    PG Test             &&&       &       &       &  \\
    $(\chi^2/\rm{d.o.f.})_{\rm{app}}$ &&&77.3 / 2 & 77.3 / 2 & 77.4 / 3 & 77.4 / 3 \\
   % $d.o.f._{\rm{app}}$    &&&  2    &   2   &   3   &    3\\    
   $(\chi^2/\rm{d.o.f.})_{\rm{dis}}$ &&& 355.8 / 3 & 535.5 / 3 & 355.8 / 4 & 535.4 / 4\\
   % $d.o.f._{\rm{dis}}$    &&&  3    &  3    &   4    &  4\\   
    $(\chi^2/\rm{d.o.f.})_{\rm{glob}}$ &&& 458.4 / 3 & 640.8 / 3 & 449.9 / 4 & 631.8 / 4\\
    %$d.o.f._{\rm{glob}}$   &&&  3    &   3   &   4    & 4 \\ 
    $(\chi^2/\rm{d.o.f.})_{\rm{PG}}$ &&& 25.2 / 2 & 28.0 / 2 & 16.7 / 3 & 19.0 / 3 \\
   % $d.o.f._{\rm{PG}}$     &&&  2    &   2   &   3    &  3 \\  
    $p$-value           &&&3.4E-6& 8.1E-7&8.0E-4&2.7E-4  \\ 
    $N \sigma$          &&&  4.5  &  4.8  &  3.2  & 3.5 \\
    \hline
  \end{tabular}
   \caption{\textbf{\textit{Summary of $\chi^2$ obtained in this global fit for different models, datasets and splits.}} Each column corresponds to a particular combination of model (3$\nu$, 3+1, or 3+1+decay) and dataset (short-baseline (SBL) or short-baseline plus IceCube, SBL+IC). The first section gives the best-fit $\chi^2$, degrees of freedom (d.o.f.), and p-value. The second section compares a given sterile model (3+1 or 3+1+decay) and the three-neutrino model via the $\chi^2$ difference for best-fit parameters for a fixed dataset. The third section contains the comparison between the 3+1 and 3+1+decay models, both with and without IceCube data. Finally, the fourth section reports the p-value for the parameter goodness of fit test for the 3+1 and 3+1+decay models, both with and without IceCube data.}
  \label{table:chi2_summary}
 \end{table*}
 
\pagebreak
\section{Oscillation probabilities at best-fit parameter points}

\begin{figure}[ht]
  \includegraphics[width=\textwidth]{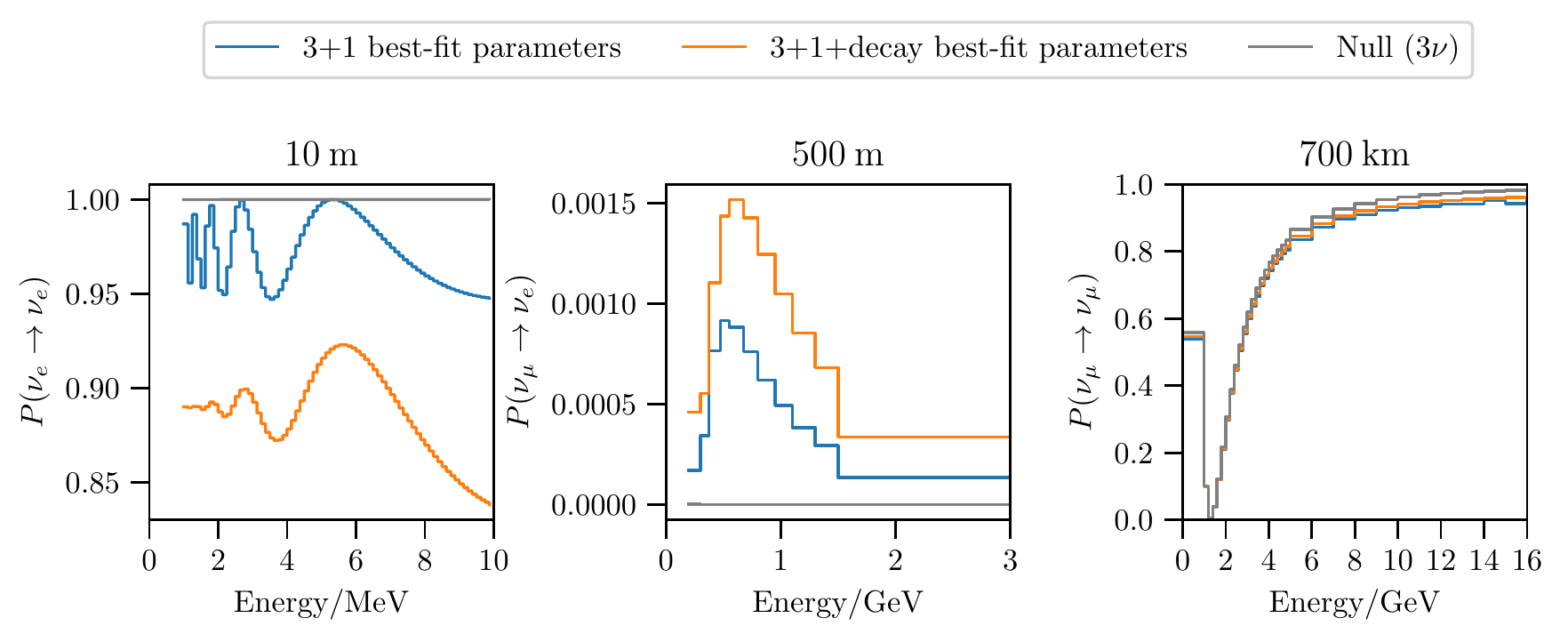}
  \caption{\textbf{\textit{Illustration of transition probabilities.}}
  In these figures, blue indicates the 3+1 model at best-fit parameters, orange indicates the corresponding point in the 3+1+decay model, and grey indicates the null hypothesis (three neutrinos).
  In each panel, probabilities plotted are the average over each bin in energy, where the binning is chosen to reflect the experimental binning. The left panel shows the electron-neutrino survival probability in the energy range relevant for reactor experiments at a ten meter baseline.
  The center panel shows the muon-neutrino to electron-neutrino appearance probability for the energy and baseline scales relevant for the MiniBooNE measurement.
  Finally, the right panel shows the muon-neutrino survival probability for the energy range of long-baseline neutrino experiments that probe the atmospheric oscillation scale, like MINOS. In this panel, the 3+1 oscillations occur too rapidly to be resolved.}
  \label{fig:oscillation_probabilities_plots}
\end{figure}

Suppl.~Fig.~\ref{fig:oscillation_probabilities_plots} shows the oscillation probabilities at the best-fit parameter points for the 3+1 and 3+1+decay models.
This figure is meant to illustrate and provide intuition on the difference between these two models and help understand why one is preferred over the other in our fits.
Thus, we compare the oscillation probabilities relevant to three categories of experiments included in our fits: (left) short-baseline reactor measurements, (center) MiniBooNE, and (right) long-baseline accelerator measurements, like MINOS. In each plot, we show the oscillation probability averaged over each energy bin, with a binning chosen to reflect that used in the relevant experiment  or experiment category.
At the longest baselines, the 3+1 oscillations vary so rapidly with energy that the energy resolution of the detector prohibits resolving individual oscillation peaks.
This makes the long-baseline muon-neutrino disappearance analysis a normalization only search, which are known to be less sensitive than shape analyses. The best-fit 3+1+decay parameters predicts a normalization slightly closer to the null hypothesis than the best-fit 3+1 parameters do.
At the smallest baselines and lowest energies of reactor experiments, the oscillations are damped and only noticeable at larger energies where the reactor neutrino statistics are smaller.
Finally, for MiniBooNE's energies and baseline, the shape predicted by the two models remains similar, but the normalization is increased in the decay scenario.
~

\end{document}